\documentclass[a4paper,fleqn,usenatbib]{mnras}

\usepackage[T1]{fontenc}
\usepackage{ae,aecompl}

\usepackage{graphicx}	
\usepackage{amsmath}	
\usepackage{amssymb}	

\title[Cosmological Tests with the FSRQ GLF]{Cosmological Tests with the FSRQ Gamma-ray Luminosity Function}

\author[H. Zeng, F. Melia \& L. Zhang]{
Houdun Zeng,$^{1,4}$\thanks{E-mail: zhd@pmo.ac.cn}
Fulvio Melia,$^{1,2}$\thanks{E-mail: fmelia@email.arizona.edu}
and Li Zhang$^{3,4}$\thanks{E-mail: lizhang@ynu.edu.cn}
\\ \\
$^{1}$Purple Mountain Observatory, Chinese Academy of Sciences, Nanjing 210008, China\\
$^{2}$Department of Physics, The Applied Math Program, and Department of Astronomy,
The University of Arizona, AZ 85721, USA\\
$^{3}$Department of Astronomy, Yunnan University, Kunming, 650091, China\\
$^{4}$Key Laboratory of Astroparticle Physics of Yunnan Province, Kunming, 650091, China
}

\date{Accepted XXX. Received YYY; in original form ZZZ}

\pubyear{2015}

\begin{document}
\label{firstpage}
\pagerange{\pageref{firstpage}--\pageref{lastpage}}
\maketitle

\begin{abstract}
The extensive catalog of $\gamma$-ray selected flat-spectrum radio quasars (FSRQs) produced
by \emph{Fermi} during a four-year survey has generated considerable interest in determining
their $\gamma$-ray luminosity function (GLF) and its evolution with cosmic time. In this paper, we
introduce the novel idea of using this extensive database to test the differential volume
expansion rate predicted by two specific models, the concordance $\Lambda$CDM and
$R_{\rm h}=ct$ cosmologies.  For this purpose, we use two well-studied formulations of
the GLF, one based on pure luminosity evolution (PLE) and the other on a luminosity-dependent
density evolution (LDDE). Using a Kolmogorov-Smirnov test on one-parameter cumulative
distributions (in luminosity, redshift, photon index and source count), we confirm the results of earlier
works showing that these data somewhat favour LDDE over PLE; we show that this is the case
for both $\Lambda$CDM and $R_{\rm h}=ct$. Regardless of which GLF one chooses, however,
we also show that model selection tools very strongly favour $R_{\rm h}=ct$ over $\Lambda$CDM.
We suggest that such population studies, though featuring a strong evolution in redshift, may
nonetheless be used as a valuable independent check of other model comparisons based solely on
geometric considerations.
\end{abstract}




\section{Introduction}

The discovery of quasars at redshifts $\gtrsim 6$ \citep{Fan2003,Jiang2007,Willott2007,Jiang2008,
Willott2010a,Mortlock2011,Venemans2013,Banados2014,Wu2015} suggests that $\gtrsim 10^{9-10}\;M_\odot$
supermassive black holes emerged only $\sim 900$ Myr after the big bang, and only $\sim 500$ Myr beyond
the formation of Population II and Population III stars \citep{Melia2013a}. Such large aggregates of
matter constitute an enduring mystery in astronomy because these quasars could not have formed so quickly
in $\Lambda$CDM without an anomalously high accretion rate \citep{Volonteri06} and/or the creation
of unusually massive seeds \citep{Yoo2004}; neither of these has actually ever been observed.
For example, \cite{Willott2010b} have recently demonstrated that no known high-$z$ quasar
accretes at more than $\sim 1-2$ times the Eddington rate (see Figure~5 in their paper; see also
Melia 2014).

This paper will feature two specific cosmologies---the aforementioned $\Lambda$CDM
(the `standard,' or concordance model) and another Friedmann-Robertson-Walker solution
known as the $R_{\rm h}=ct$ Universe \citep{Melia2007,Melia2012c,Melia2016}. Our focus
will be to explain the luminosity function of these quasars, particularly as they evolve towards
lower redshifts. Part of the motivation for this comparative study is that, unlike $\Lambda$CDM,
the $R_{\rm h}=ct$ model does not suffer from the time compression problem alluded to above
\citep{Melia2013a}. In this cosmology, cosmic reionization (starting with the creation of
Population III stars) lasted from $t\sim 883$ Myr to $\sim 2$ Gyr ($6\lesssim z\lesssim 15$),
so $\sim 5-20\;M_\odot$ black hole seeds formed (presumably during supernova explosions)
shortly after reionization had begun, would have evolved into $\sim 10^{10}\;
M_\odot$ quasars by $z\sim 6-7$ simply via the standard Eddington-limited
accretion rate. The $R_{\rm h}=ct$ Universe has thus far passed all such tests
based on a broad range of cosmological observations, but already, this consistency
with the age-redshift relationship implied by the early evolution of supermassive
black holes suggests that an optimization of the quasar luminosity function might
serve as an additional powerful discriminator between these two competing
expansion scenarios. The class of flat-spectrum radio quasars (FSRQs)
is ideally suited for this purpose.

FSRQs are bright active galactic nuclei (AGNs) that belong to a subcategory
of Blazars. These represent the most extreme class of AGNs, whose radiation
towards Earth is dominated by the emission in a relativistic jet closely aligned
with our line-of-sight. The discovery of $\gamma$-ray emission from these
sources was an important confirmation of the prediction by
\cite{Melia1989} that the particle dynamics in these jets ought to
be associated with significant high-energy emission along small viewing
angles with respect to the jet axis. It is still an open question
exactly what powers the jet activity, but it is thought that the incipient
energy is probably extracted from the black hole's spin, and is perhaps also
related to the accretion luminosity. Major mergers might have enhanced the
black-hole growth rate and activity, which would have occurred more frequently
in the early Universe. In this context, the Blazar evolution may be connected
with the cosmic evolution of the black-hole spin distribution, jet activity and
major merger events themselves, all of which may be studied via the luminosity
function (LF) and its evolution with redshift.

Recently, \emph{the Fermi Gamma-ray Space Telescope} has detected hundreds
of blazars from low redshifts out to z = 3.1, thanks to its high sensitivity \citep{Abdo10}.
Based on the previous analysis of the FSRQ $\gamma$-ray luminosity function (GLF),
it is already clear that the GLF evolution is positive up to a redshift cut-off that
depends on the luminosity (see, e.g., Padovani et al. 2007; Ajello et al. 2009;
Ajello et al. 2012). But all previous work with this sample ignored a very
important ingredient to this discussion---the impact on the GLF evolution
with redshift from the assumed cosmological expansion itself. Our main
goal in this paper is to carry out a comparative analysis of the standard
$\Lambda$CDM and $R_h = ct$ models using the most up-to-date sample of
408 FSRQs detected by the \emph{Fermi}-LAT over its four-year survey. We wish
to examine the influence on the results due to the assumed background
cosmology and, more importantly, we wish to demonstrate that the current
sample of $\gamma$-ray emitting FSRQs is already large enough for us to carry out
meaningful cosmological testing. Throughout this paper, we will be directly
comparing the flat $\Lambda$CDM cosmology with $\Omega_{\rm m} = 0.315$ and
$H_0=67.3$ km s$^{-1}$ Mpc$^{-1}$, based on the latest \emph{Planck}
results \citep{Planck14}, and the $R_{\rm h}=ct$ Universe, whose sole
parameter---the Hubble constant---will for simplicity be assumed to have the
same value as that in $\Lambda$CDM. We will demonstrate that these
data already emphatically favour $R_{\rm h}=ct$ over $\Lambda$CDM,
even without an optimization of $H_0$ for $R_{\rm h}=ct$.

The outline of this paper is as follows. In \S~2, we will summarize the observational
data, specifically the 3FGL catalog \citep{Acero15}, and describe how the $\gamma$-ray
luminosity is determined for each specific model. \S~3 will provide an account of
the critical differences between these two cosmologies that directly impact the calculation
of the GLF, and we discuss the currently preferred ansatz for this luminosity function
based on the most recent analysis of these data in \S~4. We present and discuss our
results in \S~5, and conclude in \S~6.

\section{Observational Data and Source Sample}

The third Fermi Large Area Telescope source catalog (3FGL) provided by \cite{Acero15}
lists 3,303 sources detected by \emph{Fermi}-LAT during its four years of operation.
These data include the source location and its spectral properties. A subset of these
is the third LAT AGN catalog (3LAC; Ackermann et al. 2015), containing 1,591 AGNs of
various types located at high galactic latiude, i.e., $\mid b \mid \geq 10^o$. Most of
the detected AGNs are blazars, which consist of 467 FSRQs, 632 BL Lacs, 460 blazar
candidates of uncertain type (BCUs), and 32 non-blazar AGNs. Removing the entries
in 3LAC for which the corresponding $\gamma$-ray sources were not associated with AGNs,
had more than one counterpart or were flagged for other reasons in the analysis,
\cite{Ackermann15} reduced the AGN catalog to a `clean' sample of 1,444 sources,
including 414 FSRQs, 604 BL Lacs, 402 BCUs and 24 non-blazar AGNs. The energy flux
distribution of all the \emph{Fermi} sources may be seen in Figure~18 of
\cite{Acero15}. The flux threshold in 3FGL is $\simeq 3 \times 10^{-12}$ erg cm$^{-2}$
s$^{-1}$, lower than the value ($\simeq 5 \times 10^{-12}$ erg cm$^{-2}$ s$^{-1}$)
in 2FGL and ($\simeq 8 \times 10^{-12}$ erg cm$^{-2}$ s$^{-1}$) in 1FGL. The
sample above the 3FGL flux threshold is essentially complete (see Figure~18 of Acero
et al. 2015). Note also that all of the FSRQs cataloged in 3FGL have measured redshifts.
In this paper, we have chosen to use only the FSRQs from 3LAC, and not
the BL Lacs, because of their greater redshift coverage and better sample completeness,
both of which strengthen our statistical analysis.

Our sample is therefore comprised of the 414 FSRQs detected by \emph{Fermi} with a test
statistic (TS) $\geq 25$ and latitude $|b|\geq 10^o$. Their $\gamma$-ray fluxes $S_\gamma$
and photon indices $\Gamma$ in the energy range 0.1 GeV - 100 GeV are obtained from
3FGL\footnote{URL: $http://fermi.gsfc.nasa.gov/ssc/data/access/lat/4yr_catalog/$} and
their redshifts are from Table 7 of \cite{Ackermann15}.\footnote{URL: $http://www.asdc.asi.it/fermi3lac/$}
We calculate the $\gamma$-ray luminosity $L_{\gamma}$ of an FSRQ using the expression:
\begin{equation}
\label{eq:L} L_{\gamma}=\frac{4\pi D^2_L(z)S_{\gamma}}{1+z}K,
\end{equation}
where $D_L(z)$ is the luminosity distance at redshift $z$ and $K$ is the $K-$correction for the
observed fluxes using $S_{\gamma} = S_{\gamma,\rm obs}(1+z)^{\Gamma-1}$.  The
corresponding photon flux $F_{\gamma}$ (in units of photons cm$^{-2}$ s$^{-1}$) in the 0.1
GeV - 100 GeV energy band used in this paper is readily obtained from the following
expressions (see, e.g., Ghisellini et al. 2009, Singal et al. 2014):
\begin{eqnarray}
\label{eq:S-F}
F_{\gamma}=
 \left\{
\begin{array}{lcl}
S_{\gamma} \frac{2-\Gamma}{1-\Gamma} \frac{1}{E_{0.1}}\frac{1-10^{3(1-\Gamma)}}{1-10^{3(2-\Gamma)}} &
& {\rm (if~~~ \Gamma \neq 2.0)} \\
S_{\gamma} \frac{1}{1-\Gamma} \frac{E_{0.1}^{1-\Gamma}}{\ln(10^3)} (10^{3(1-\Gamma)}-1) &
& {\rm (if~~~\Gamma = 2.0)}\;,
\end{array}
\right.
\end{eqnarray}
where $E_{0.1}=1.602 \times 10^{-4}$ erg (corresponding to 0.1 GeV) is the lower energy limit.

\begin{figure*}
  \begin{center}
  \begin{tabular}{cc}
\hspace{-1cm}
	\includegraphics[width=\columnwidth]{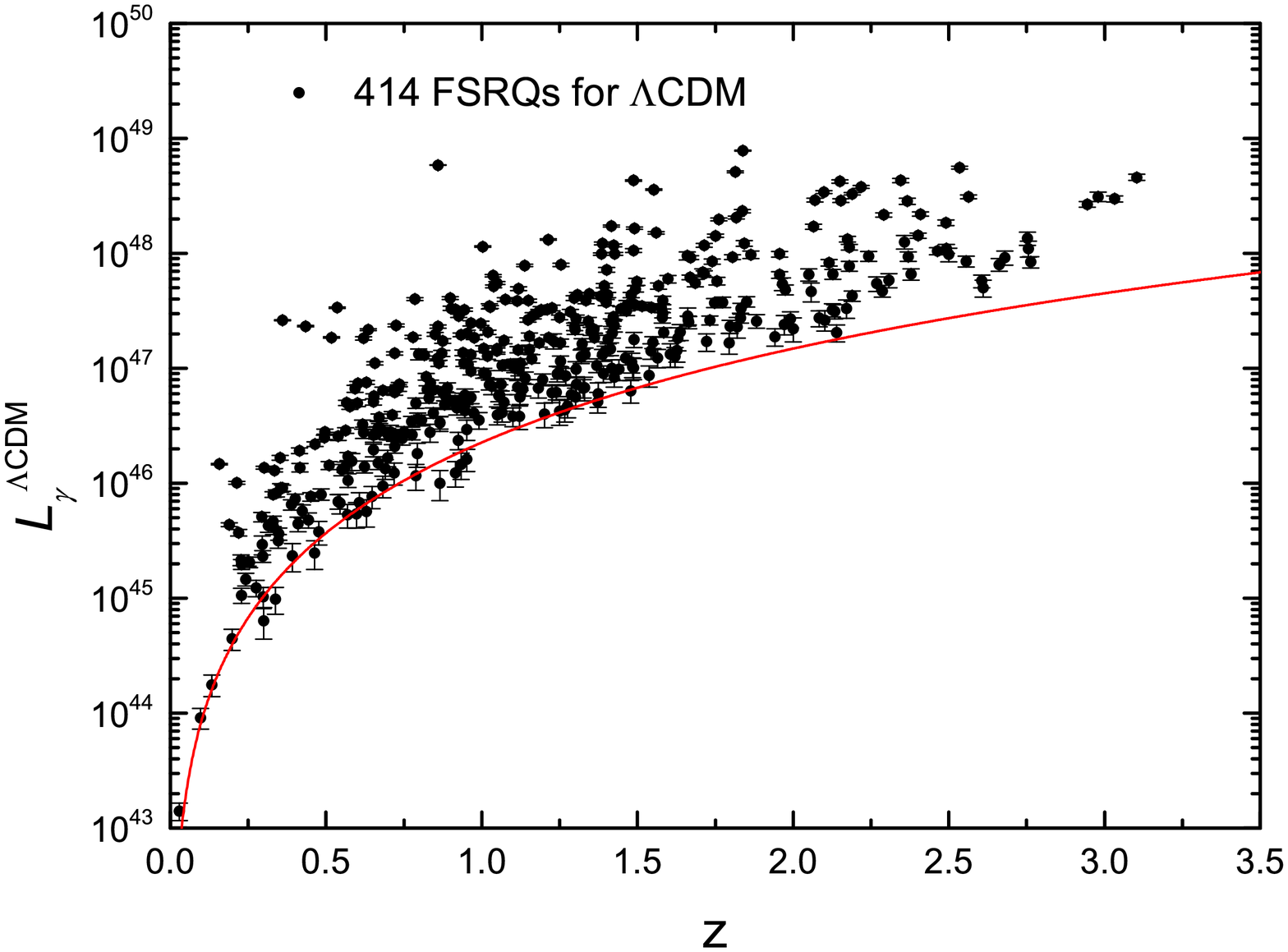} &
\hspace{0cm}
    \includegraphics[width=\columnwidth]{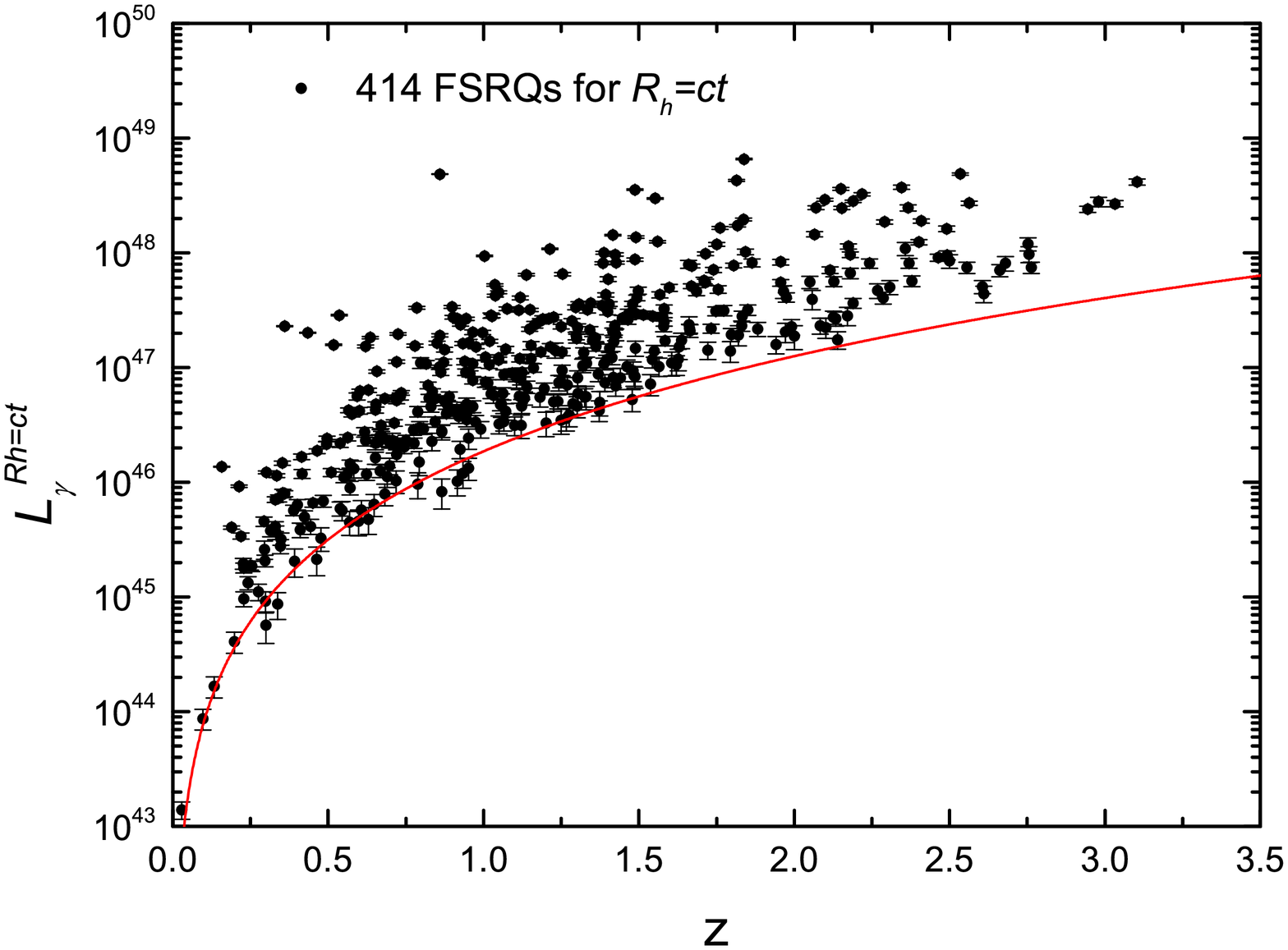}
    \end{tabular}
  \end{center}
    \caption{The 3LAC FSRQ luminosity-redshift distribution for both $\Lambda$CDM
(left panel) and $R_{\rm h}=ct$ (right panel). The solid curves are calculated
using Equation~(\ref{eq:L}) for the threshold flux $S_{\gamma, \rm limit}= 3.0 \times 10^{-12}$ erg cm$^{-2}$ s$^{-1}$, and a fixed photon index $\Gamma =2.44$, which is the mean of the $\Gamma$ distribution from 3LAC \citep{Ackermann15}.}
    \label{fig:figure1}
\end{figure*}

Clearly, a determination of $L_\gamma$ requires the assumption of a particular cosmological
model. A detailed description of the differences between the two models we are considering
here, the concordance $\Lambda$CDM cosmology and the $R_{\rm h}=ct$ Universe, may be
found in \cite{Melia2012a,Melia2012b,Melia2013a,Melia2013b}. The quantity most relevant
to the analysis in this paper is the luminosity distance $D_L(z)$, which in $\Lambda$CDM is
given as
\begin{eqnarray}
\label{eq:Dl-Gamma}
D_{L}^{\Lambda{\rm CDM}}(z)&\hskip-0.1in=\hskip-0.1in&\frac{c}{H_0} \frac{(1+z)}{\sqrt{|\Omega_k|}}
 {\rm sinn}\Bigg\{\mid\Omega_{k}\mid^{1/2}  \times \nonumber \\
 &\hskip -0.1in\null\hskip-0.1in&\hskip-0.3in\left.\int_0^{z} \frac{dz}{\sqrt{(1+z)^2(1+\Omega_m z)-
z(2+z) \Omega_{\Lambda}}} \right\} ,
\end{eqnarray}
where $c$ is the speed of light and $H_0$ is the Hubble constant at
the present time. In this equation, $\Omega_{\rm m} \equiv \rho_{\rm m}/\rho_{\rm c}$ and
$\Omega_{\Lambda} \equiv \rho_{\Lambda}/\rho_{\rm c}$ are, respectively,
the energy density of matter and dark energy written in terms of today's critical
density ($\rho_{\rm c}\equiv 3c^2H_0^2/8\pi G$), and $\Omega_k$ is the
spatial curvature of the Universe, appearing as a term proportional to the spatial
curvature constant k in the Friedmann equation. Also, $sinn$ is $\sinh$ when
$\Omega_k > 0$ and $\sin$ when $\Omega_k < 0$. For a flat universe with
$\Omega_k = 0$, Equation~(\ref{eq:Dl-Gamma}) simplifies to the form $(1 + z)c/H_0$
times the integral. In the $R_{\rm h} = ct$ Universe, the luminosity distance is given
by the much simpler expression
\begin{equation}
\label{eq:Dl-R}
D_{L}^{R_{\rm h}=ct}(z)= R_{\rm h}(t_0)(1+z)\ln(1+z)\,,
\end{equation}
where $R_{\rm h}(t_0) = c/H_0$ is the gravitational horizon (equal to the Hubble radius)
at the present time.

The luminosities in these two models are simply related according to the expression
\begin{equation}
L_{\gamma}^{R_{\rm h}=ct}(z) = L_{\gamma}^{\Lambda{\rm CDM}}(z) \times \left(\frac{D_{L}^{R_{\rm h}=ct}(z)}
{D_{L}^{\Lambda{\rm CDM}}(z)}\right)^2\;.
\end{equation}
Figure~\ref{fig:figure1} shows the resulting luminosity-redshift distribution for both $\Lambda$CDM
(left panel) and $R_{\rm h}=ct$ (right panel). In this figure, the solid curves are calculated
using Equation~(\ref{eq:L}) for the threshold flux $S_{\gamma, \rm limit}= 3.0 \times 10^{-12}$ erg cm$^{-2}$ s$^{-1}$, and a fixed photon index $\Gamma =2.44$, which is the mean of the $\Gamma$ distribution from 3LAC \citep{Ackermann15}.
For a given observed flux, $S_{\gamma}$, the inferred luminosity depends on the
assumed cosmology through the model-dependent luminosity distance, $D_L$, as indicated in
Equation~(5). The data points in figure~1 are therefore slightly different for the two models
being compared here. However, since we are assuming the concordance model with {\it Planck}
parameters (see below), it turns out that the ratio $D_{L}^{\Lambda{\rm CDM}}/D_{L}^{R_{\rm h}=ct}$
is very close to $1$ over the redshift range $0\lesssim z\lesssim 3$. One may see this in figure~3
of Melia (2015), which plots this ratio for several values of the matter density $\Omega_m$.
For the {\it Planck} matter density $\Omega_m= 0.308 +/- 0.012$, $D_L^{\Lambda{\rm CDM}}$
is just a few percent bigger than $D_{L}^{R_{\rm h}=ct}$ over the entire redshift range considered
in this paper.

\section{Model Comparison}

Given a fixed background cosmology, we constrain the model parameters of the FSRQs GLF (see \S~4)
using the method of maximum likelihood evaluation (see, e.g., Chiang \& Mukherjee 1998; Ajello et al. 2009;
Norumoto et al. 2006; Abdo et al. 2010b; Zeng et al. 2014). The likelihood function $\mathcal{L}$ is defined
by the expression
\begin{equation}
\label{eq:Likelihood}
\mathcal{L} = \textrm{exp}(-N_{\rm{exp}})\; \Pi^{N_{\rm obs}}_{i=1} \frac{d^3 N(z_i,L_{\gamma,i},\Gamma_i)}{dz\,dL_\gamma\,d\Gamma},
\end{equation}
where $d^3N/dz\,dL_\gamma\,d\Gamma$ is the space density of FSRQs, which generally depends on the luminosity
function, $\rho(z,L_\gamma)$; the intrinsic distribution of photon indices with a
Gaussian dependence, $dN/d\Gamma \propto \rm{exp}(-\frac{(\Gamma-\mu)^2}{2\sigma^2})$, where
$\mu$ and $\sigma$ are the Gaussian mean and dispersion, respectively; and the comoving
volume element per unit redshift and unit solid angle, $dV_{\rm com}/dz$. This space density may
be expressed as
\begin{eqnarray}
\label{eq:SD}
\frac{d^3 N}{dz\,dL_{\gamma}\,d\Gamma} &=&  \frac{d^2 N}{dL_\gamma\,dV_{\rm com}} \times \frac{dN}{d\Gamma} \times
\frac{dV_{com}}{dz} \nonumber \\
&=& \rho(z,L_\gamma) \times \frac{dN}{d\Gamma} \times
\frac{dV_{\rm com}}{dz}\;.
\end{eqnarray}
In Equation~(6), the quantity $N_{\rm exp}$ is the expected number of FSRQ $\gamma$-ray detections,
\begin{eqnarray}
\label{eq:N_exp}
N_{\rm exp} &=&  \int^{\Gamma_{\rm max}}_{\Gamma_{\rm min}}
\int^{z_{\rm max}}_{z_{\rm min}} \int^{L_{\gamma,\rm max}}_{L_{\gamma,\rm min}}
\frac{d^3 N}{dz\,dL_{\gamma}\,d\Gamma} \times \nonumber \\
&\null& \qquad\qquad\qquad\qquad\quad\omega(F_{\gamma}, \Gamma) dz\,d\Gamma d\,L_{\gamma}\;.
\end{eqnarray}
Based on the properties of the $\gamma$-ray emitting FSRQ sample, we take $\Gamma_{\rm min}=1.0$,
$\Gamma_{\rm max}=3.0$, $z_{\rm min}=0.0$, $z_{\rm max}=6.0$,  $L_{\gamma,\rm min}=
1.0\times 10^{43}$ erg s$^{-1}$, and $L_{\gamma,\rm max}=1.0\times 10^{52}$ erg s$^{-1}$.
The quantity $\omega(F_{\gamma}, \Gamma)$ is the detection efficiency, and
represents the probability of detecting a FSRQ with the photon flux $F_{\gamma}$
and photon index $\Gamma$ (e.g., Atwood et al. 2009; Abdo et al. 2010c; Ajello et al.
2012; Zeng et al. 2013; DiMauro et al. 2014a, 2014b), where $F_{\gamma}$ is strongly
dependent on the photon index $\Gamma$ (obtained from Equations~1 and 2), and is
also a function of the luminosity $L_{\gamma}$ and redshift $z$. To estimate
$\omega(F_{\gamma},\Gamma)$, we use the method provided by Di Mauro et al.
(2014a). The efficiency $\omega(F_{\gamma}, \Gamma)$ at a photon flux $F_{\gamma}^k \in
(F_{\gamma}^{k,min}, F_{\gamma}^{k,max})(k={1 \cdot \cdot \cdot \cdot N_1})$ and
photon index $\Gamma^l \in (\Gamma^{l,min}, \Gamma^{l,max}) (l={1 \cdot \cdot
\cdot \cdot N_2})$ may be estimated as
\begin{equation}
\omega(F_{\gamma}, \Gamma) = (1+\eta) \frac{N_{\rm sources}^{k,l}}{\Delta \Omega
\int_{F_{\gamma}^{k,min}}^{F_{\gamma}^{k,max}} \frac{dN}{dF_{\gamma}} dF_{\gamma}
\int_{\Gamma^{l,min}}^{\Gamma^{l,max}} \frac{dN}{d\Gamma} d\Gamma}\;,
\label{eq:efficieny}
\end{equation}
where $\Delta \Omega$ is the solid angle associated with $|b| > 10^o$, $\eta$
is the incompleteness of the sample representing the ratio of unassociated sources to the
total number of sources, and $N_{\rm sources}^{k,l}$ is the number of selected sources.
The integrated values of $dN/dF_{\gamma}$ and $dN/d\Gamma$ are, respectively, the intrinsic
flux and index distributions of the sources. Note that the effect of the photon index
on the detection efficiency is fully accounted for in our method, the details of which
are discussed in the Appendix of Di Mauro et al. (2014a). Here we assume that the intrinsic
flux distributions in the low flux band of FSRQs and blazars have the same power-law index,
because the distribution log$N$-log$S$ of FSRQs is flatter than that of blazars at low fluxes
(see Figure 14 of Abdo et al. 2010c).
Figure. 11 shows the efficiency for this sample of FSRQs evaluated
using Eq. \ref{eq:efficieny}, in the region $\Gamma \in [1.5,3.2]$, compared with the
1FGL \citep{Abdo10c} and 2FGL \citep{DiMauro2014a} samples. This comparison shows
that this method of evaluation is a simplified and effective way to find the efficiency, though
the correct method would include a simulation and an estimate of the number of detected
sources versus the number of simulated objects in different flux bins \citep{Abdo10c,Ackermann16}.

While maximization of the likelihood function $\mathcal{L}$ is an appropriate and reliable method
for optimizing the parameters of a given model, to determine statistically which of the models
is actually preferred by the data it is now common in cosmology to use several model selection
tools (see, e.g., Melia \& Maier 2013, and references cited therein). These include the Akaike
Information Criterion, ${\rm AIC}\equiv-2\ln \mathcal{L}+2n$, where $n$ is the number of
free parameters (Liddle 2007); the Kullback Information Criterion, ${\rm KIC}=-2\ln \mathcal{L}+3n$
(Cavanaugh 2004); and the Bayes Information Criterion, ${\rm BIC}=-2\ln \mathcal{L}+(\ln N_{\rm obs})n$,
where $N_{\rm obs}$ is the number of data points (Schwarz 1978).
A more quantitative ranking of models can be computed as follows. When using the AIC,
with ${\rm AIC}_\alpha$ characterizing model $\mathcal{M}_\alpha$, the
unnormalized confidence in $\mathcal{M}_1$ is given by the Akaike weight exp$(- \rm {AIC}_1/2)$.
The relative probability that $\mathcal{M}_1$ is statistically preferred is
\begin{equation}
P(\mathcal{M}_1) = \frac{\rm {exp}(-AIC_1/2)}{\rm {exp}(-AIC_1/2)+\rm {exp}(-AIC_1/2)}\;.
\end{equation}
The difference $\Delta_{\rm AIC} \equiv {\rm AIC}_2\nobreak-{\rm AIC}_1$ determines the
extent to which $\mathcal{M}_1$ is favoured over~$\mathcal{M}_2$. For Kullback and
Bayes, the likelihoods are defined analogously. In using these model selection tools,
the outcome $\Delta_{\rm AIC}$ (and analogously for KIC and BIC) is judged to represent
`positive' evidence that model 1 is to be preferred over model 2 if $\Delta_{\rm AIC}> 2.0$.
If $2 < \Delta_{\rm AIC} <6$, the evidence favouring model 1 is moderate, and it is
very strong when $\Delta_{\rm AIC}>10$.

\begin{figure}
	\includegraphics[width=\columnwidth]{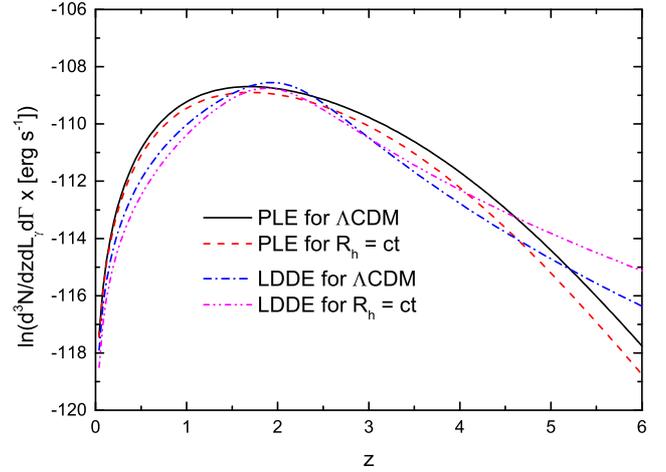}
    \caption{Space density of $\gamma$-ray emitting FSRQs (Equation~7) as a function of $z$ for
the concordance $\Lambda$CDM and $R_{\rm h}=ct$ cosmologies, using two formulations
of the GLF (described in \S~4): PLE (pure luminosity evolution) and LDDE (luminosity-dependent
density evolution). This illustration corresponds to two fixed parameters:
$L_{\gamma}=1.0\times 10^{48}$ erg s$^{-1}$ and $\Gamma =2.44$.}
    \label{fig:figure2}
\end{figure}

In this paper, we have chosen to compare two specific models: the concordance $\Lambda$CDM
cosmology, with Planck-measured prior values for all its parameters, and the $R_{\rm h}=ct$
Universe, whose sole parameter---the Hubble constant $H_0$---is, for simplicity, assumed to
have the same value as that in $\Lambda$CDM. The total number of free parameters in
this study is therefore limited in both cases to the formulation of the $\gamma$-ray luminosity
function (GLF) (see \S~4), which is common to both models. In other words, the model
selection statistic we will be using, i.e., $\Delta$ (for AIC, KIC, or BIC, as the case may
be), depends solely on the quantity $W\equiv -2\ln\mathcal{L}$ and in fact, for this reason,
all three of these information criteria share the same values of $\Delta$. As it turns out,
$W$ represents the $\chi^2$ distribution. Transforming to the standard expression, we may
therefore write
\begin{equation}
W=-2 \,\Sigma_i^{N_{\rm obs}}\;\ln \frac{d^3 N}{dz\,dL_{\gamma}\,d\Gamma} +2 N_{\rm exp}\;.
\end{equation}

For each model, we optimize the GLF parameters using the Markov Chain Monte Carlo (MCMC)
technology, which is widely applied to give multidimensional parameter constraints from
observational data. In practice, this
means we will find the parameter values that minimize $W$, which yields the best-fit parameters
and their associated $1\sigma$ errors. We have adapted the MCMC code from \texttt{COSRAYMC}
\citep{Liu2012}, which itself was adapted from the \texttt{COSMOMC} package \citep{Lewis2002}.
Additional details about the MCMC method may be found in \cite{Gamerman1997} and
\cite{Mackay2003}.  We remark here that this way of posing the maximum
likelihood problem is different from that chosen by \cite{Ajello09,Ajello12,Ajello14} but, as
pointed out in \cite{Ajello09}, who tested these various approaches, one gets exactly
the same results using these different formulations, so there is no preference for one
over the other, except in terms of convenience.

\begin{figure}
	\includegraphics[width=\columnwidth]{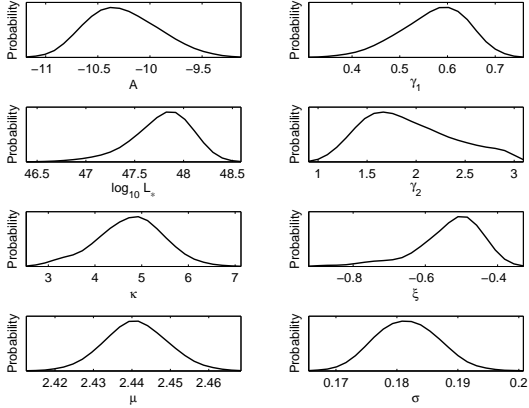}
    \caption{The 1D probability distributions of the GLF parameters and their 1 $\sigma$ statistical
uncertainties for the concordance $\Lambda$CDM cosmology, assuming PLE.}
    \label{fig:figure3}
\end{figure}

One of the principal differences between the two models affecting the value of
$W$ is the space density of FSRQs due to its dependence on the luminosity distance
$D_L(z)$. The comoving differential volume is given as
\begin{equation}
\label{eq:com-v}
\frac{dV_{\rm com}}{dz} = D_{\rm com}^2 \frac{dD_{\rm com}}{dz}\;,
\end{equation}
where $D_{\rm com}\equiv D_L/(1+z)$ is the comoving distance. For flat $\Lambda$CDM
and $R_{\rm h}=ct$ we have, respectively,
\begin{equation}
\frac{dD_{\rm com}^{\Lambda{\rm CDM}}}{dz} = \frac{c}{H_0} \frac{1}{\sqrt{(1+z)^2(1+\Omega_{\rm m} z)-
z(2+z) \Omega_{\Lambda}}}\;,
\end{equation}
and
\begin{equation}
\frac{dD_{\rm com}^{R_{\rm h}=ct}}{dz} = \frac{c}{H_0} \frac{1}{1+z}.\;
\end{equation}
For illustration, we show in Figure~\ref{fig:figure2} the space density (Equation~7) as a function of redshift $z$,
using the two formulations of the GLF described below, i.e., for pure luminosity evolution (PLE) and
for luminosity-dependent density evolution (LDDE). The choice of parameters in this example is
based on the discussion in \S~5.

\section{The Gamma-ray Luminosity Function}
The so-called pure luminosity evolution (PLE) formulation for the GLF is motivated
by the observed space density of radio-quiet AGNs, peaking at intermediate
redshifts that correlate with source luminosity (see, e.g., Ueda et al. 2003;
Hasinger et al. 2005). This peak may correspond to the combined effect of
black-hole growth and the falloff in fueling activity. The conventional formulation
for the space density of this GLF (see, e.g., Ajello et al. 2012) has the form
\begin{eqnarray}
\label{eq:GPLE}
\rho(L_{\gamma},z)&\hskip-0.1in=\hskip-0.1in&\rho(L_{\gamma}/e[z]) \nonumber \\
&\hskip-0.1in=\hskip-0.1in&\frac{A\,e(z)}{\ln10\;L_{\gamma}}\left[
\left(\frac{L_{\gamma}/e(z)}{L_{\ast}}\right)^{\gamma_1}\hskip-0.1in+\left(\frac{L_{\gamma}/e(z)}
{L_{\ast}}\right)^{\gamma_2}\right],\quad
\end{eqnarray}
where $e(z) = (1+z)^{\kappa}e^{z/\xi}$ is the evolution factor correlated with source
luminosity, $A$ is a normalization factor, $L_{\ast}$ is the evolving break luminosity,
$\gamma_1$ is the faint-end slope index, $\gamma_2$ is the bright-end slope index,
and $\kappa$ and $\xi$ represent the redshift evolution. Including the additional
parameters $\mu$ and $\sigma$ characterizing the (Gaussian) photon index
distribution (see discussion following Equation~6) then results in a total
of 8 parameters that need to be optimized in our PLE analysis.

However, while the PLE GLF generally provides a good fit to the observed redshift
and luminosity distributions, it is a very poor representation of the observed
$\log N$-$\log S$ (Ajello et al. 2012). Closer scrutiny of the values of $\kappa$
and $\xi$ in in different redshift bins suggests that there is a significant shift
in the redshift peak, with the low- and high-luminosity samples peaking at
$\sim 1.15$ and $\sim 1.77$, respectively.

\begin{figure}
	\includegraphics[width=\columnwidth]{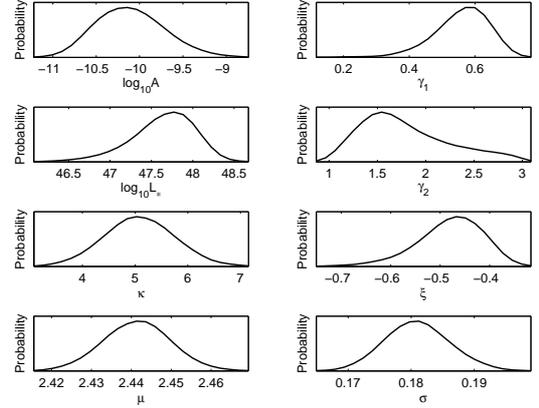}
    \caption{The 1D probability distributions of the GLF parameters and their 1 $\sigma$ statistical
uncertainties for the $R_{\rm h}=ct$ cosmology, assuming PLE.}
    \label{fig:figure4}
\end{figure}

Since the simple PLE GLF may not be a completely adequate fit to the \emph{Fermi} data,
and since the redshift peak apparently evolves with luminosity, it is also beneficial to
consider a GLF with luminosity-dependent density evolution (LDDE; see Ueda et al. 2003;
Ajello et al. 2012). In this formulation, the GLF evolution is decided by a redshift cut-off
that depends on luminosity. The space density for this GLF is given by the expression
\begin{eqnarray}
  \label{eq:GLDDE}
  \rho(L_{\gamma},z)&\hskip-0.1in=\hskip-0.1in&\frac{A}{\ln10\;L_{\gamma}}
  \left[\left(\frac{L_{\gamma}}{L_{\ast}}\right)^{\gamma_1}+
  \left(\frac{L_{\gamma}}{L_{\ast}}\right)^{\gamma_2}\right]^{-1}\times \nonumber \\
  &\hskip-0.1in\null\hskip-0.1in&\left[\left(\frac{1+z}{1+z_c(L_{\gamma})}\right)^{p_1}+
  \left(\frac{1+z}{1+z_c(L_{\gamma})}\right)^{p_2}\right]^{-1},\qquad
\end{eqnarray}
with
\begin{equation}
z_c(L_{\gamma})\equiv z_{c}^{\ast}(L_{\gamma}/10^{48})^{\alpha}\;,
\end{equation}
where $A$ is a normalization factor, $L_{\ast}$ is the evolving break luminosity, $\gamma_1$ and $p_1$
are the faint-end slope indeces, $\gamma_2$ and $p_2$ are the bright-end slope indeces,
$z_{c}^{\ast}$ is the redshift peak with luminosity $10^{48}$ ergs s$^{-1}$,
and $\alpha$ is the power-law index of the redshift-peak evolution.

Previous studies based solely on the concordance $\Lambda$CDM model (see, e.g.,
Ajello et al. 2012) have shown that the LDDE provides a good fit to the LAT data and
can reproduce the observed distribution quite well. The log-likelihood ratio test
strongly favours it over PLE. In this paper, we will use both formulations of the
GLF, just to be sure that we are not biasing our results prematurely with an
ansatz for $ \rho(L_{\gamma},z)$ that is too specific. As it turns out, both
the PLE and LDDE formulations give completely consistent results when it
comes to model selection. We will therefore conclude that the results of our
model comparison using the $\gamma$-ray emitting FSRQs is not at all
dependent on assumptions concerning the form of the GLF.

\begin{table*}
	\centering
    \tiny
	\caption{Optimized parameters (with 1 $\sigma$ errors) of the PLE GLF for $\Lambda$CDM and $R_{\rm h}=ct$.}
	\label{tab:table1}
	\begin{tabular}{llllllllllll} 
		\hline\hline
		PLE & $H_0^{a}$ & $\Omega_{\rm m}$ &
log$_{10}A^{b}$ & $\gamma_1$ &
log$_{10} L_{\ast}$ & $\gamma_2$  & $\kappa$&
$\xi$ & $\mu$ &
$\sigma$ &
$W=$\\
&&&&&&&&&&&$-2\ln\mathcal{L}$\\
		\hline
		$\Lambda$CDM &67.3 &0.315 &$-10.24^{+0.36}_{-0.34}$ & $0.57^{+0.06}_{-0.06}$ & $47.75^{+0.31}_{-0.31}$ & $1.90^{+0.53}_{-0.47}$ & $4.76^{+0.63}_{-0.62}$ & $-0.53^{+0.07}_{-0.07}$ & $2.44^{+0.01}_{-0.01}$ &
$0.18^{+0.01}_{-0.01}$& 89106\\
       $R_{\rm h} = ct$ &67.3 & &$-10.11^{+0.40}_{-0.39}$ & $0.56^{+0.08}_{-0.08}$ & $47.64^{+0.36}_{-0.37}$ & $1.81^{+0.52}_{-0.46}$ & $5.08^{+0.64}_{-0.65}$ & $-0.48^{+0.06}_{-0.06}$ & $2.44^{+0.01}_{-0.01}$ &
       $0.18^{+0.01}_{-0.01}$ & 88962 \\
        \hline\hline
       KS Test&$P_D(L_\gamma)$ &$P_D(z)$&$P_D(\Gamma)$ \\
        \hline
	   $\Lambda$CDM & 98.4\% &67.5 \% &72.9 \%  \\
       $R_{\rm h} = ct$ & 98.6\% &73.4 \% &74.9 \%\\
		\hline
     \multicolumn{3}{l}{$^a$ In units of km s$^{-1}$ Mpc$^{-1}$}\\
\multicolumn{3}{l}{$^b$ In units of Mpc$^{-3}$ erg$^{-1}$ s}\\
	\end{tabular}
\end{table*}

\section{Results and Discussion}
\subsection{Pure Luminosity Evolution}
We begin with the PLE assumption and optimize the GLF parameters by maximizing the
likelihood function $\mathcal{L}$ using the MCMC method. The best-fit parameters and
one-dimensional (1D) probability distributions are shown in Figure~\ref{fig:figure3} for $\Lambda$CDM
and Figure~\ref{fig:figure4} for $R_{\rm h}=ct$. The mean-fit parameter values and their 1 $\sigma$
confidence levels are listed in Table \ref{tab:table1}.

\begin{figure}
	\includegraphics[width=\columnwidth]{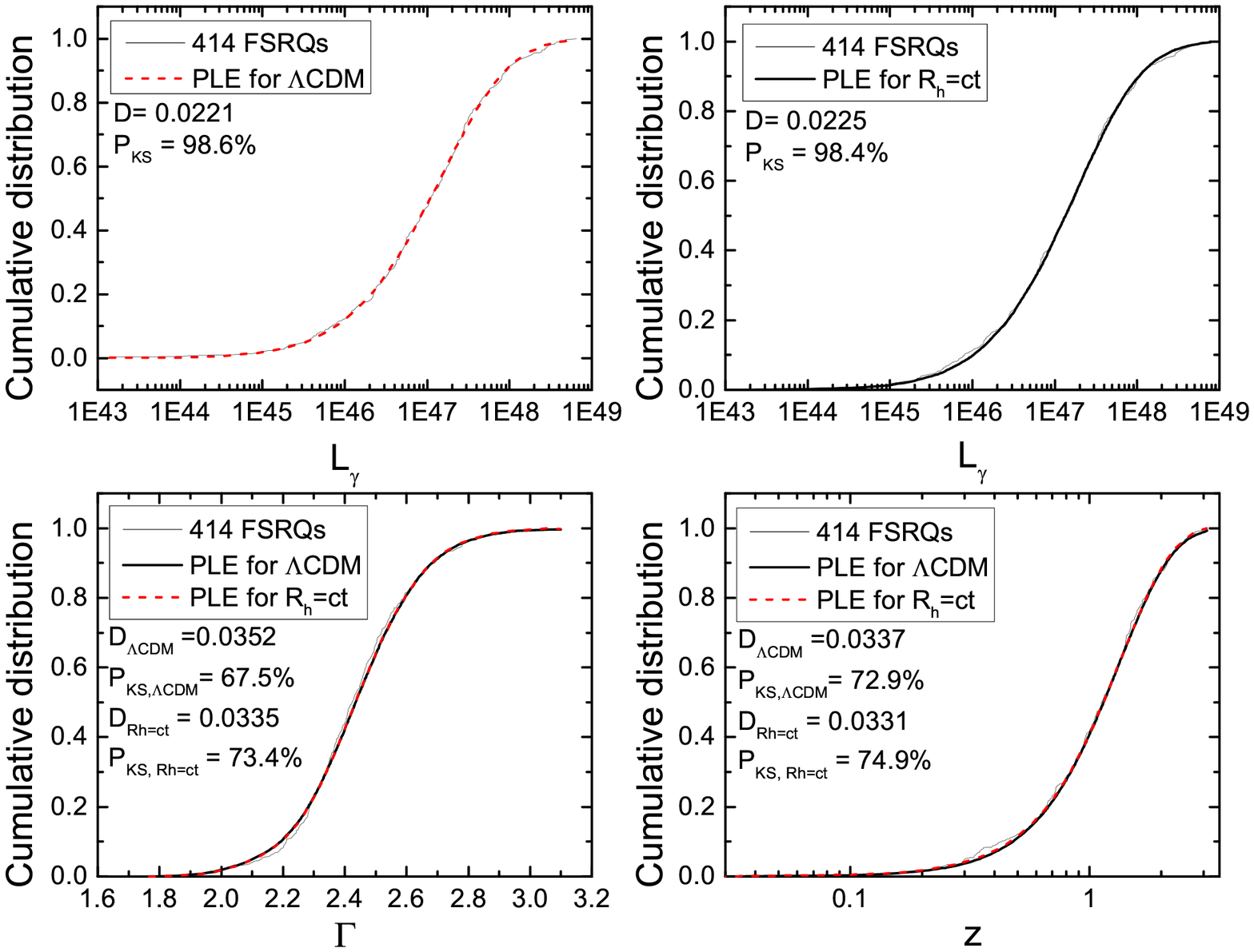} \\
    \includegraphics[width=\columnwidth]{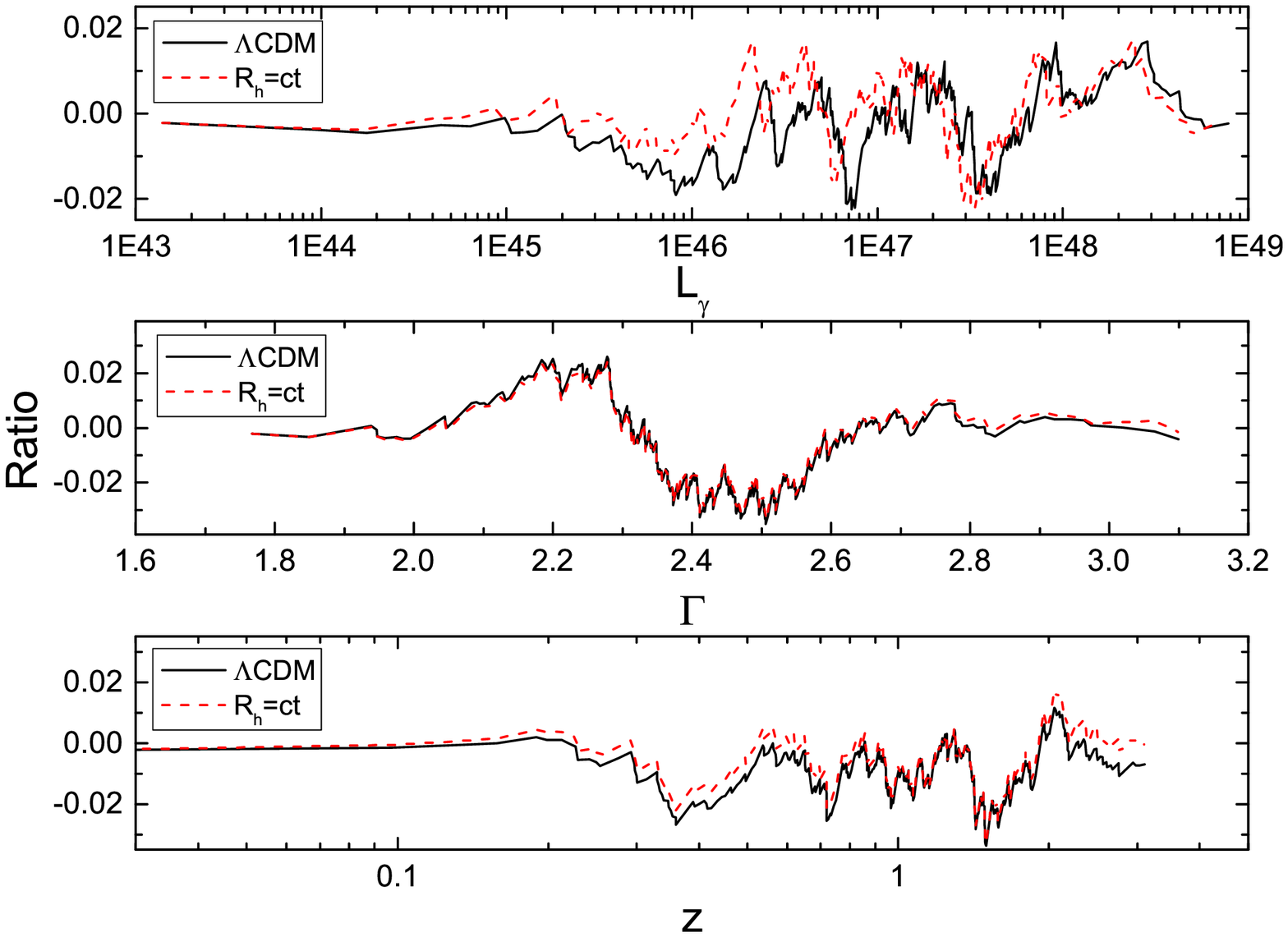}
    \caption{Top: The predicted cumulative distributions (in luminosity, redshift and photon index)
from the luminosity functions with best-fit parameters, versus the observed distributions
of the corresponding quantities, assuming a PLE GLF, for both $\Lambda$CDM and the
$R_{\rm h}=ct$ Universe. Bottom: To bring out the model differences more clearly, we
also plot in the bottom 3 panels the ratio of predicted to observed distributions for the
luminosity, redshift and photon index. The line definitions correspond to those in
the upper panels. In these lower 3 panels, the top is for luminosity, the middle is
for photon index, and the lowest is for redshift.}
    \label{fig:figure5}
\end{figure}

Ajello et al. (2012) examined whether the PLE GLF could adequately account for the
\emph{Fermi} data, though strictly only for the $\Lambda$CDM cosmology, and concluded
that it was not a good representation of the $\log N$-$\log S$ distribution. To see whether
this is still true for our sample, and also to test whether this defiiciency is also present for
the $R_{\rm h}=ct$ Universe, we apply the Kolmogorov-Smirnov (KS) test for the
predicted one-parameter cumulative distributions using the measured populations
as individual functions of redshift, luminosity and photon index. The theoretical
one-parameter distributions are calculated as follows:
\begin{eqnarray}
N(<z) &=& \int^{\Gamma_{\rm max}}_{\Gamma_{\rm min}}
\int^{z}_{z_{\rm min}} \int^{L_{\gamma, \rm max}}_{L_{\gamma, \rm min}}
\frac{d^3 N}{dz\,dL_{\gamma}\,d\Gamma}\times \nonumber \\
&&\qquad\qquad\qquad\omega(F_{\gamma}\,\Gamma)\, dz\,d\Gamma\, dL_{\gamma}\;,
\end{eqnarray}
\begin{eqnarray}
N(<L_{\gamma}) &=& \int^{\Gamma_{\rm max}}_{\Gamma_{\rm min}}
\int^{z_{\rm max}}_{z_{\rm min}} \int^{L_{\gamma}}_{L_{\gamma, \rm min}}
\frac{d^3 N}{dz\,dL_{\gamma}\,d\Gamma} \times\nonumber \\
&&\qquad\qquad\qquad\omega(F_{\gamma}\,\Gamma)\, dz\,d\Gamma\, dL_{\gamma}\;,
\end{eqnarray}
\begin{eqnarray}
N(<\Gamma) &=& \int^{\Gamma}_{\Gamma_{\rm min}}
\int^{z_{\rm max}}_{z_{\rm min}} \int^{L_{\gamma, \rm max}}_{L_{\gamma, \rm min}}
\frac{d^3 N}{dz\,dL_{\gamma}\,d\Gamma} \times\nonumber \\
&&\qquad\qquad\qquad\omega(F_{\gamma}\,\Gamma)\, dz\,d\Gamma\, dL_{\gamma}\;.
\end{eqnarray}
In addition, the source count distribution is given by the expression
\begin{eqnarray}
N(> F_{\gamma}) &\hskip-0.11in=\hskip-0.15in& \int^{\Gamma_{\rm max}}_{\Gamma_{\rm min}}
\int^{z_{\rm max}}_{z_{\rm min}} \int^{L_{\gamma, \rm max}}_{\rm{Max}(L_{\gamma}[F_{\gamma},
z, \Gamma], L_{\gamma, \rm min})}
\frac{d^3 N}{dz\,dL_{\gamma}\,d\Gamma}\times\nonumber \\
&\hskip-0.1in\null\hskip-0.1in& \qquad\qquad\qquad\qquad\qquad\omega(F_{\gamma}\,\Gamma)\, dz\,d\Gamma\, dL_{\gamma}\;,
\end{eqnarray}
where $L_{\gamma}(F_{\gamma}, z, \Gamma)$ is the luminosity of a source at redshift $z$ with photon
index $\Gamma$, having a flux $F_{\gamma}$. The corresponding curves, together with
the binned data with which they are compared, are shown in Figures~\ref{fig:figure5} and \ref{fig:figure9} for
both $\Lambda$CDM and $R_{\rm h}=ct$.

At least visually, the predicted cumulative distributions appear to match the data
quite well, aside from the source count distribution. Indeed, the PLE GLF passes the KS
test in three cumulative distributions (luminosity, redshift, photon index) for both
cosmologies, though only at a modest level of confidence in the case of $\Gamma$. The
$R_{\rm h}=ct$ Universe does better than the concordance $\Lambda$CDM model for all the
distributions. As we shall see shortly, the results of our KS comparison between the
cumulative distributions for PLE and LDDE are somewhat mixed. Certainly in the case of $z$,
the LDDE GLF passes the KS test with a significantly higher level of confidence, where it
reaches $\sim 98\%$ in the case of $R_{\rm h}=ct$, compared to only $\sim 74\%$ for PLE.
However, the KS test results for the cumulative distributions in $L_{\gamma}$ are very
similar between PLE and LDDE. Note that the left panels in Figure~\ref{fig:figure9} show that the predicted
distributions are a very poor representation of the observed log$N$-log$S$. Based solely
on the cumulative redshift distribution, together with the relatively poor source-count
distribution, we do confirm the result in Ajello et al. (2012), that the LDDE GLF
appears to be a better representation of the \emph{Fermi} data than the PLE GLF.

\subsection{Luminosity-Dependent Density Evolution}
\begin{figure}
	\includegraphics[width=\columnwidth]{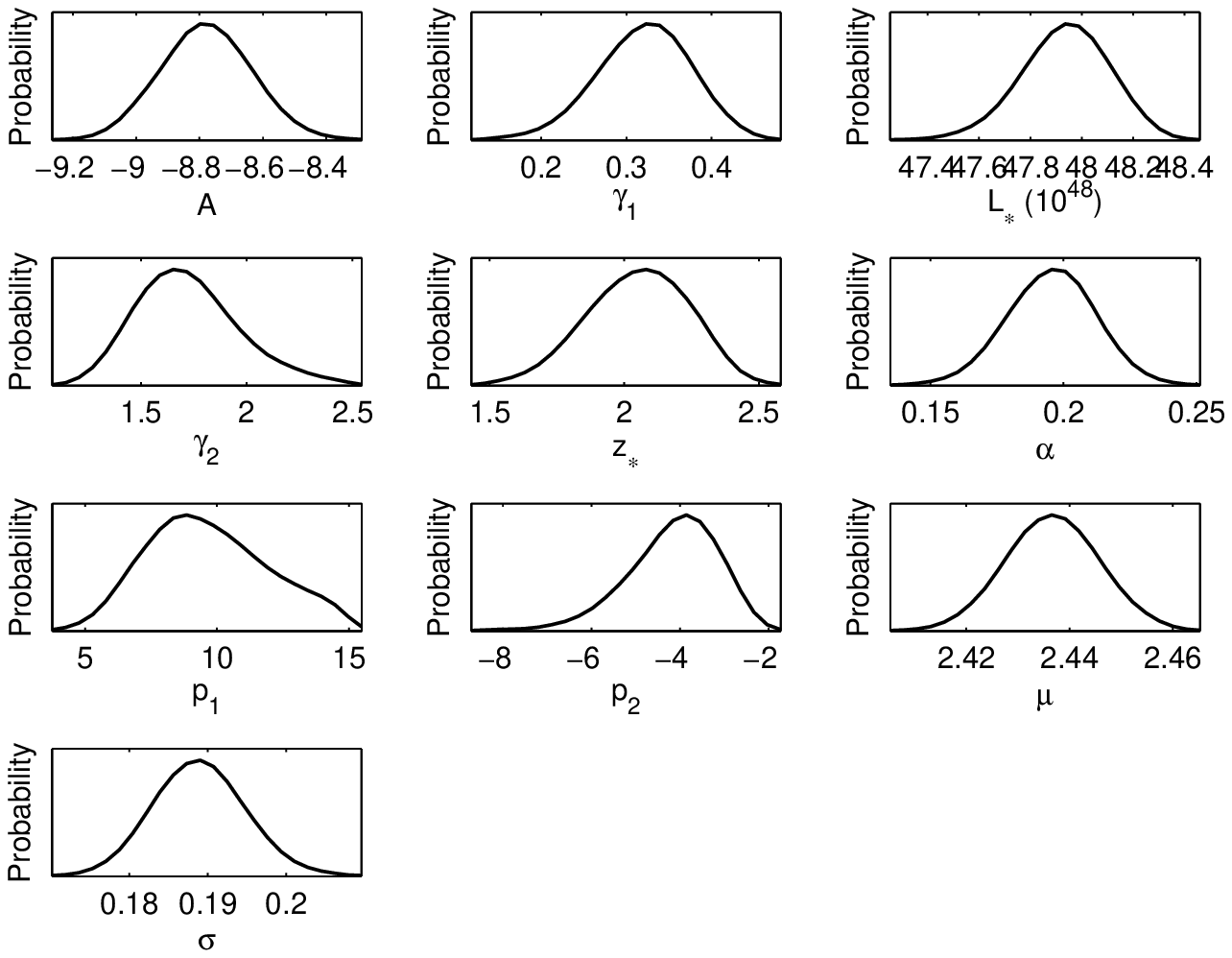}
    \caption{The 1D probability distributions of the GLF parameters and their 1 $\sigma$ statistical
uncertainties for the concordance $\Lambda$CDM cosmology, assuming LDDE.}
    \label{fig:figure6}
\end{figure}

\begin{figure}
	\includegraphics[width=\columnwidth]{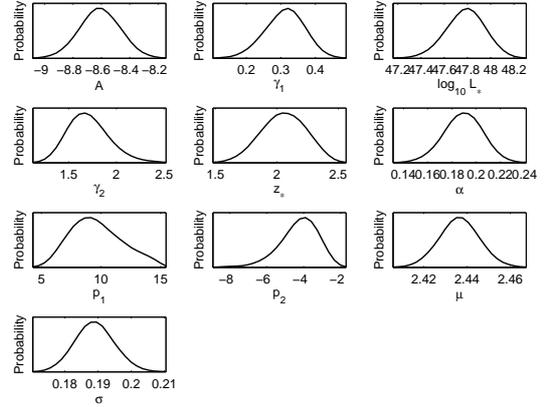}
    \caption{The 1D probability distributions of the GLF parameters and their 1 $\sigma$ statistical
uncertainties for the $R_{\rm h}=ct$ cosmology, assuming LDDE.}
    \label{fig:figure7}
\end{figure}

\begin{figure}
	\includegraphics[width=\columnwidth]{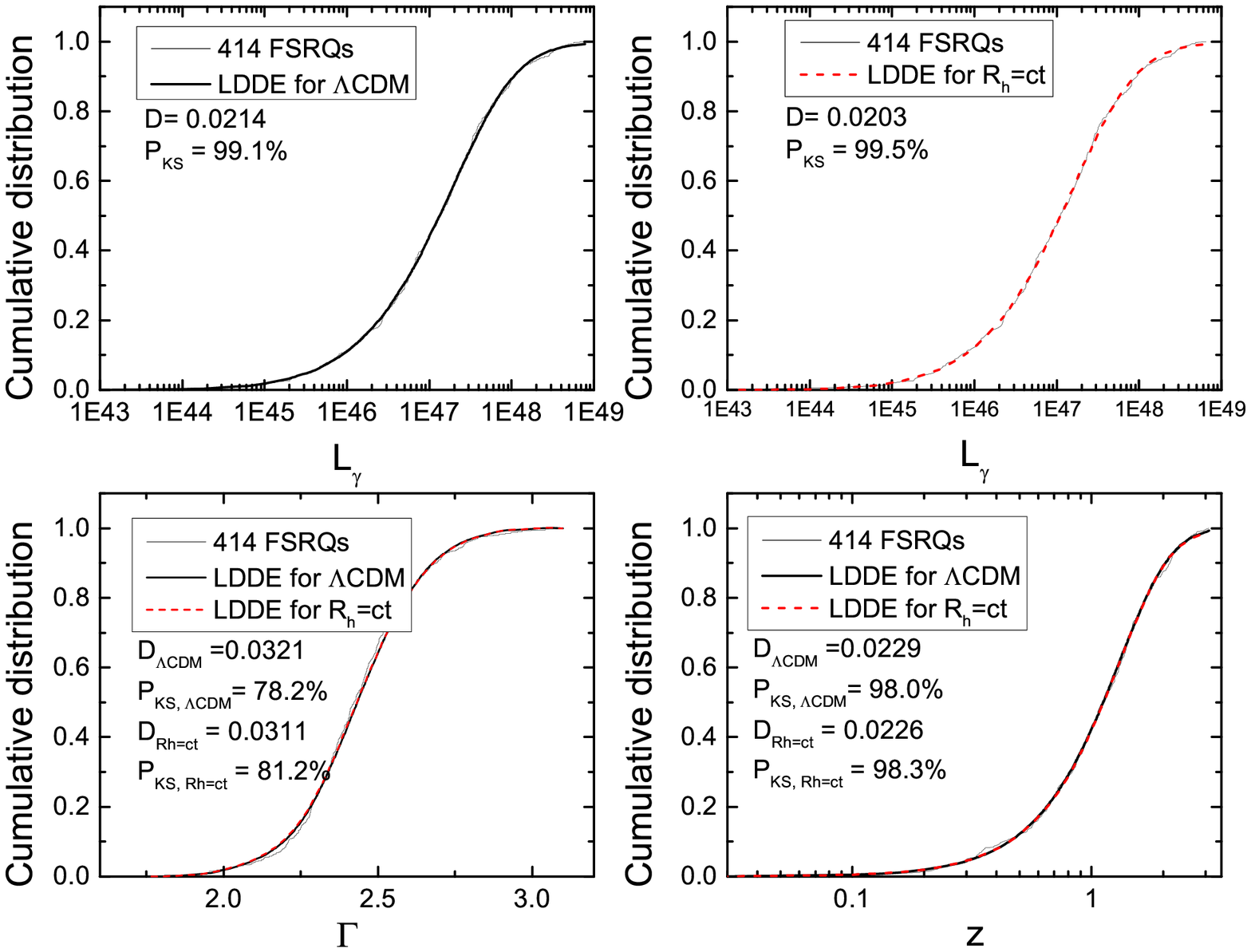} \\
    \includegraphics[width=\columnwidth]{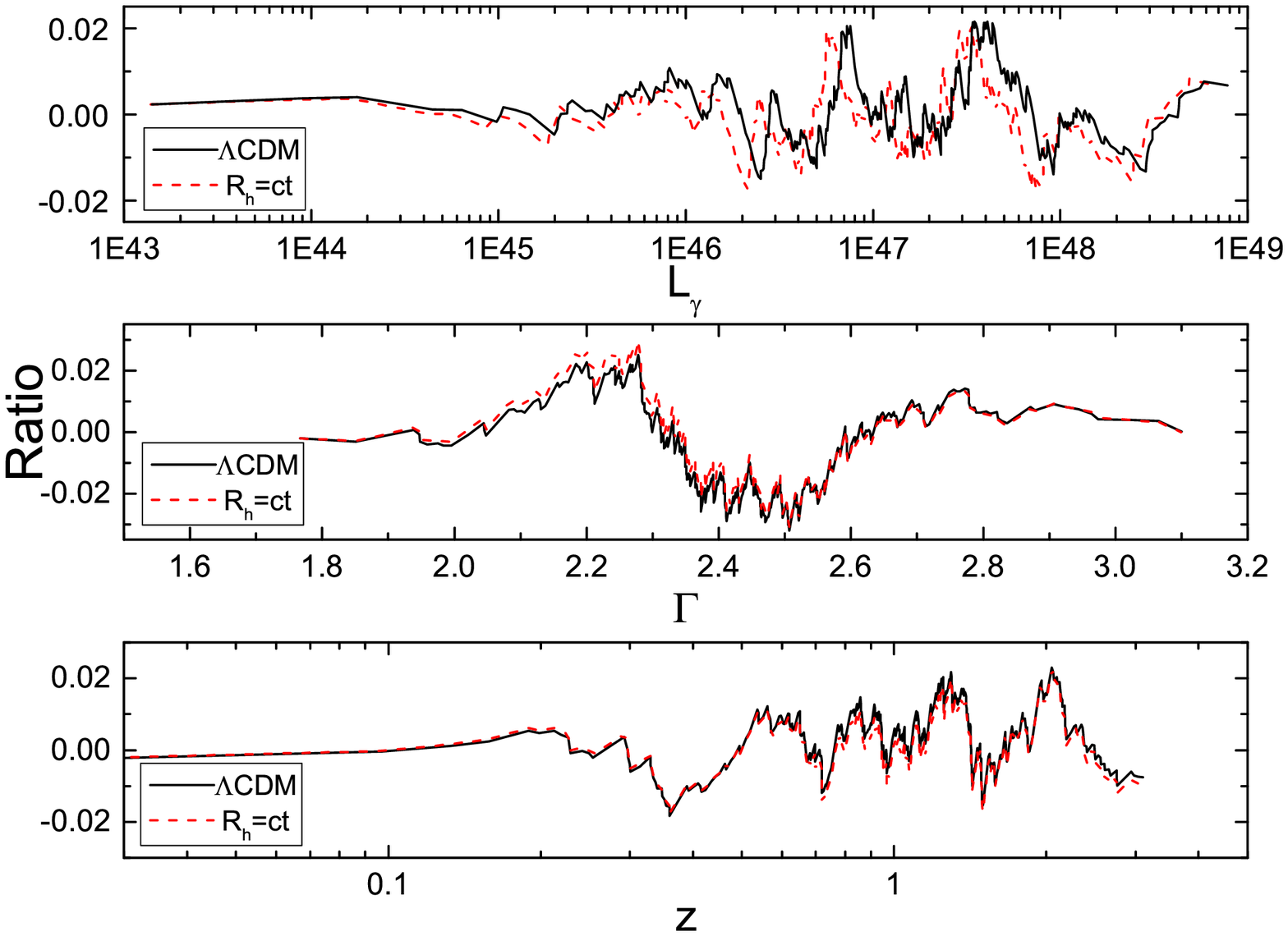}
    \caption{Top: The predicted cumulative distributions (in luminosity, redshift and photon index)
from the luminosity functions with best-fit parameters, versus the observed distributions
of the corresponding quantities for the LDDE GLF, for both $\Lambda$CDM and the $R_{\rm h}=ct$
Universe. Bottom: As in fig.~5, we also show here the ratio of predicted to observed distributions
for luminosity, redshift and photon index, corresponding the distributions in the upper panels.
 The respective confidence levels are listed in Table \ref{tab:table2}.}
    \label{fig:figure8}
\end{figure}

We next optimize the ten parameters of the LDDE GLF, given in Equation~(16), using
the MCMC method to maximize the likelihood function for the same sample of 414 FSRQs that
we used for PLE. The 1D probability distributions of these
parameters are shown in Figure~\ref{fig:figure6} for $\Lambda$CDM and Figure~\ref{fig:figure7} for $R_{\rm h}=ct$.
Their mean-fit values and 1 $\sigma$ confidence levels are listed in Table \ref{tab:table2}. As before,
and specifically to examine which GLF is a better match to the data, we carry out the
same Kolmogorov-Smirnov test as for PLE, and compare the predicted one-parameter
cumulative distributions with the data in Figure~\ref{fig:figure8}. The favorable visual impression one
gets is confirmed by the confidence levels of the matches, which are quoted in Table \ref{tab:table2}.

\begin{table*}
	\centering
    \tiny
	\caption{Optimized parameters (with 1 $\sigma$ errors) of the LDDE GLF for $\Lambda$CDM and $R_{\rm h}=ct$.}
	\label{tab:table2}
  \begin{tabular}{p{0.9cm}p{0.7cm}p{0.7cm}p{0.9cm}p{0.9cm}p{0.9cm}
           p{0.9cm}p{0.9cm}p{0.9cm}p{0.9cm}p{0.9cm}p{0.9cm}p{0.9cm}p{0.9cm}}
		\hline\hline
		LDDE & $H_0$$^a$ & $\Omega_{\rm m}$ &
 log$_{10}A^b$ & $\gamma_1$ &
log$_{10} L_{\ast}$ & $\gamma_2$ & $z_{c}^{\ast}$ &
$\alpha$ & $p_1$ &
$p_2$ & $\mu$ & $\sigma$ &
$W=$\\
&&&&&&&&&&&&&$-2\ln\mathcal{L}$\\
		\hline
		$\Lambda$CDM & $0.673$ &0.315 &$-8.78^{+0.14}_{-0.14}$ & $0.32^{+0.05}_{-0.05}$ & $47.93^{+0.17}_{-0.17}$ & $1.71^{+0.23}_{-0.23}$ & $2.06^{+0.19}_{-0.19}$ & $0.20^{+0.02}_{-0.02}$ &$9.71^{+2.53}_{-2.36}$&$-4.15^{+0.99}_{-1.00}$& $2.44^{+0.01}_{-0.01}$ &
$0.19^{+0.01}_{-0.01}$& 89047 \\[1ex]
$R_{\rm h}=ct$ &0.673 & &$-8.61^{+0.14}_{-0.14}$ & $0.31^{+0.06}_{-0.06}$ & $47.78^{+0.17}_{-0.17}$ & $1.70^{+0.21}_{-0.21}$ & $2.06^{+0.19}_{-0.18}$ & $0.19^{+0.02}_{-0.02}$ &$9.64^{+2.31}_{-2.18}$&$-4.23^{+0.99}_{-1.00}$& $2.44^{+0.01}_{-0.01}$ &
$0.19^{+0.01}_{-0.01}$ & 88903 \\[1ex]
        \hline\hline
       KS Test&$P_D(L_\gamma)$ &$P_D(z)$&$P_D(\Gamma)$ \\
        \hline
	   $\Lambda$CDM & 99.1\% &78.2 \% &98.0 \%  \\[1ex]
       $R_h = ct$ & 99.5\% &81.2 \% &98.3 \%  \\[1ex]
		\hline
    \multicolumn{3}{l}{$^a$ In units of km s$^{-1}$ Mpc$^{-1}$}\\
\multicolumn{3}{l}{$^b$ In units of Mpc$^{-3}$ erg$^{-1}$ s}\\
	\end{tabular}
\end{table*}

\begin{figure*}
  \begin{center}
  \begin{tabular}{cc}
\hspace{-1cm}
	\includegraphics[width=\columnwidth]{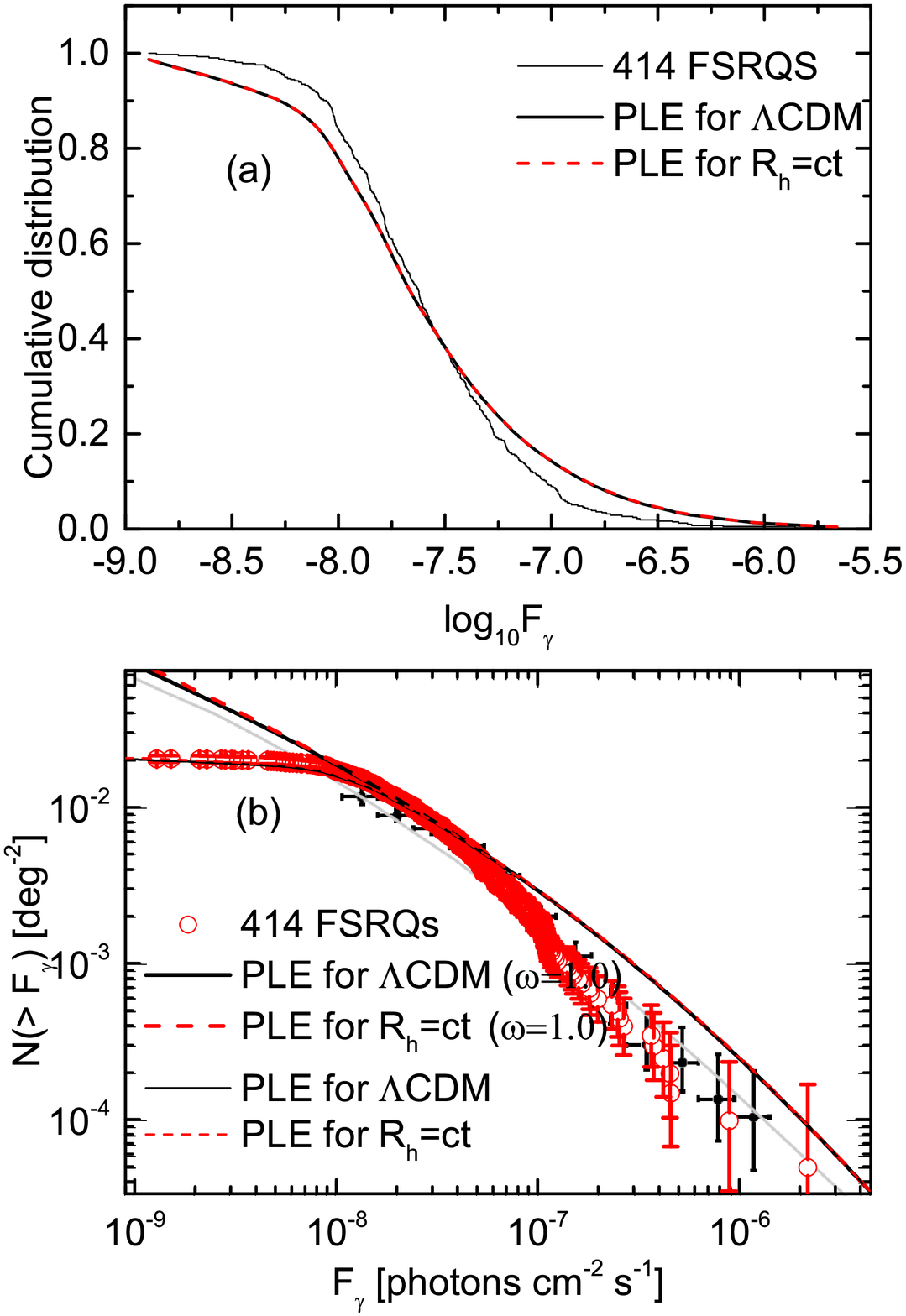} &
\hspace{0cm}
    \includegraphics[width=\columnwidth]{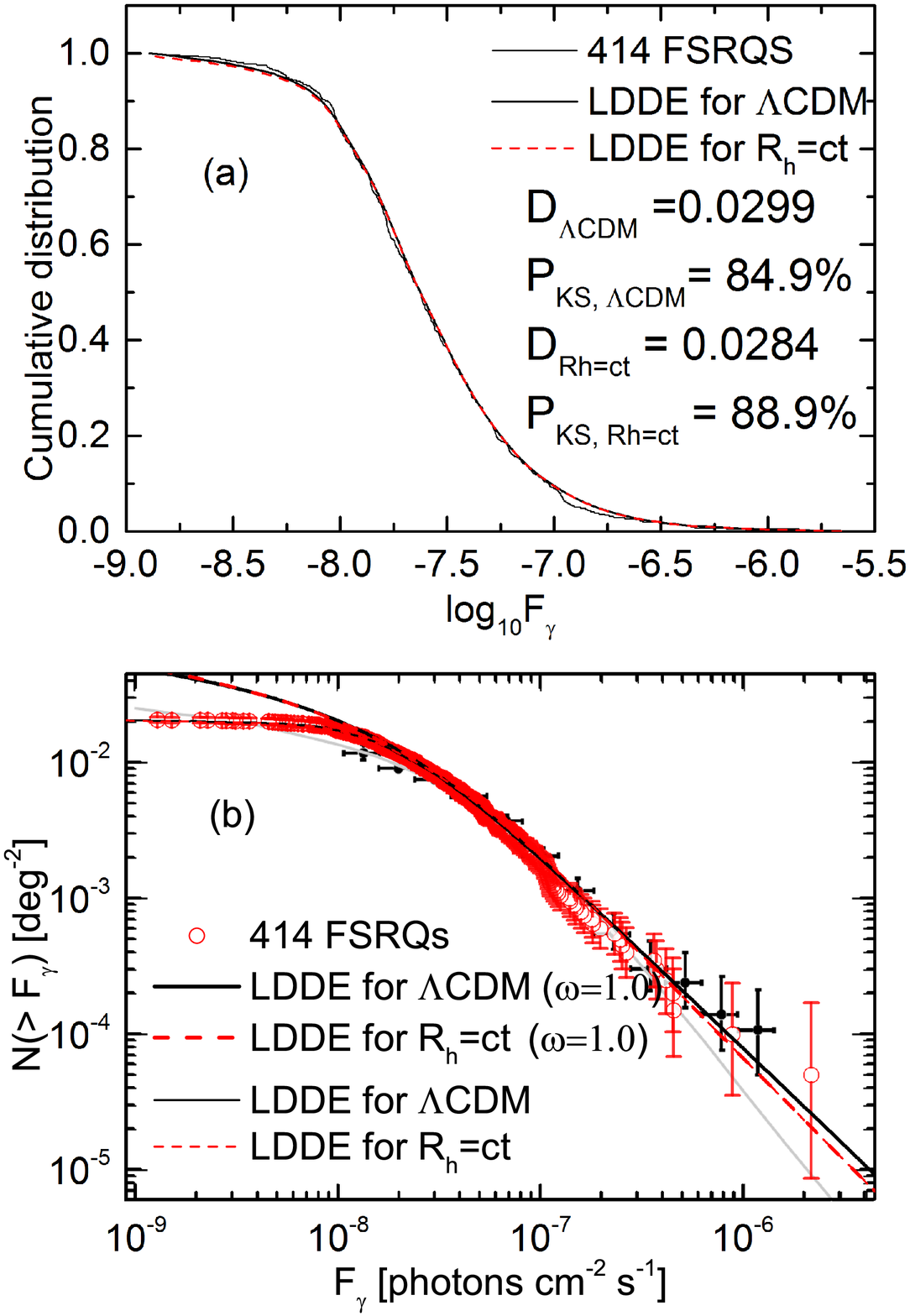}
        \end{tabular}
  \end{center}
    \caption{ The source count distribution of FSRQs for both cosmologies, assuming PLE (left panels)
and LDDE (right panels). The curves are the best-fit models reported in the text for the
$\Lambda$CDM (solid) and $R_{\rm h}=ct$ (dashed) cosmologies. (a) The cumulative distributions
in photon flux.(b) Solid circles represent the observed cumulative distribution in photon flux for the 186
FSRQs with $TS \geq 50$ and $|b|\geq 15^{o}$ reported in Ajello et al. (2012); empty circles represent the
observed cumulative distribution for the 414 FSRQs in our sample. For comparison, the (additional) thick
curves in the lower panels are the intrinsic cumulative distributions assuming
$\omega(F_{\gamma}\,\Gamma)=1.0$.}
    \label{fig:figure9}
\end{figure*}

The $R_{\rm h}=ct$ cosmology does at least as well as $\Lambda$CDM, and usually better,
in all the KS tests using the various one-parameter cumulative FSRQ distributions.
In both models, the predicted source count distribution is a better match to the data
for LDDE than for PLE (the right-hand panels of Figure~\ref{fig:figure9}), supporting the conclusion
drawn earlier by Ajello et al. (2012) that LDDE is favoured over PLE
by the measured $\log N$-$\log S_\gamma$ relation.

\begin{figure*}
  \begin{center}
  \begin{tabular}{cc}
\hspace{-1cm}
	\includegraphics[width=\columnwidth]{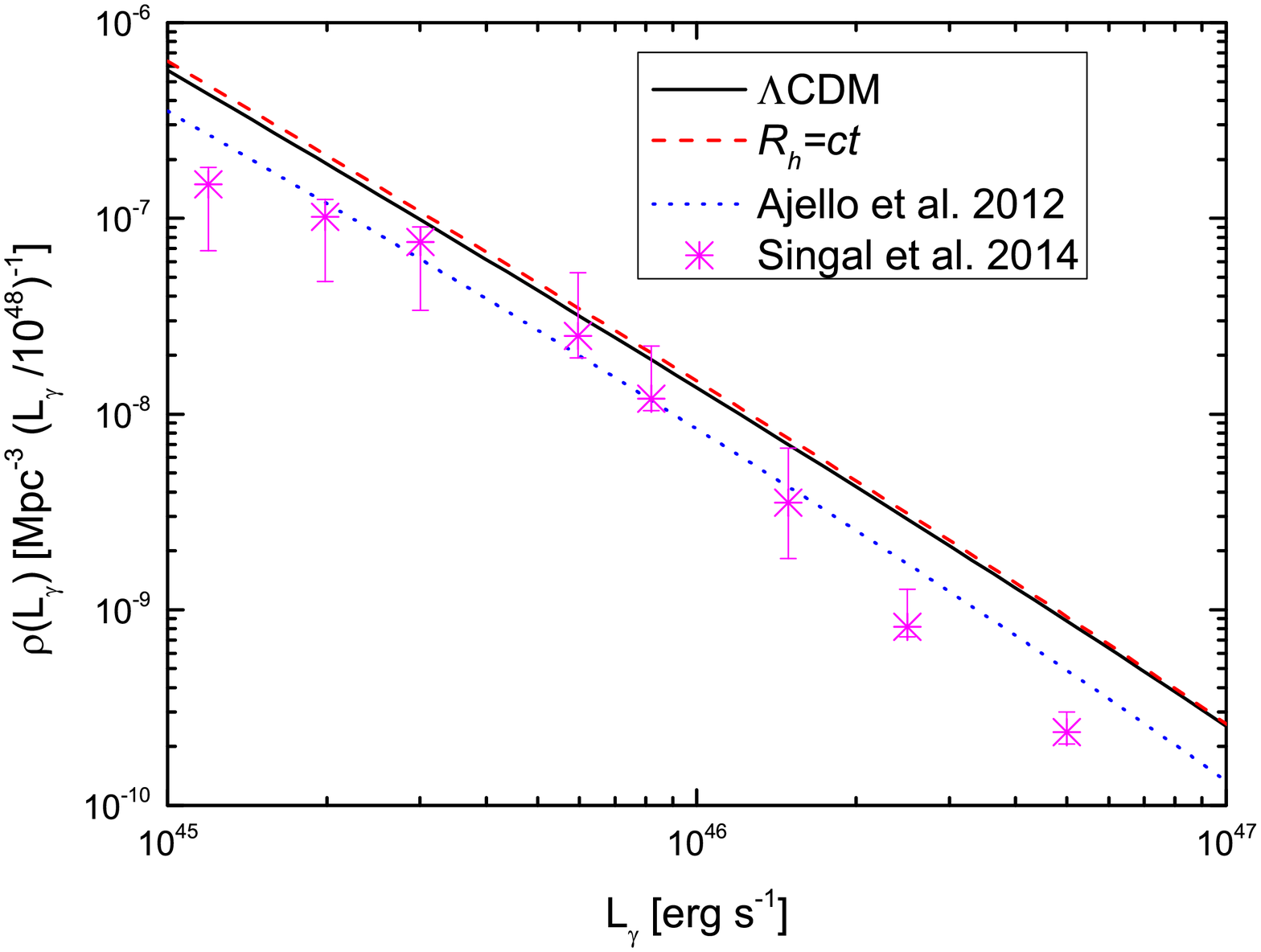} &
\hspace{0cm}
    \includegraphics[width=\columnwidth]{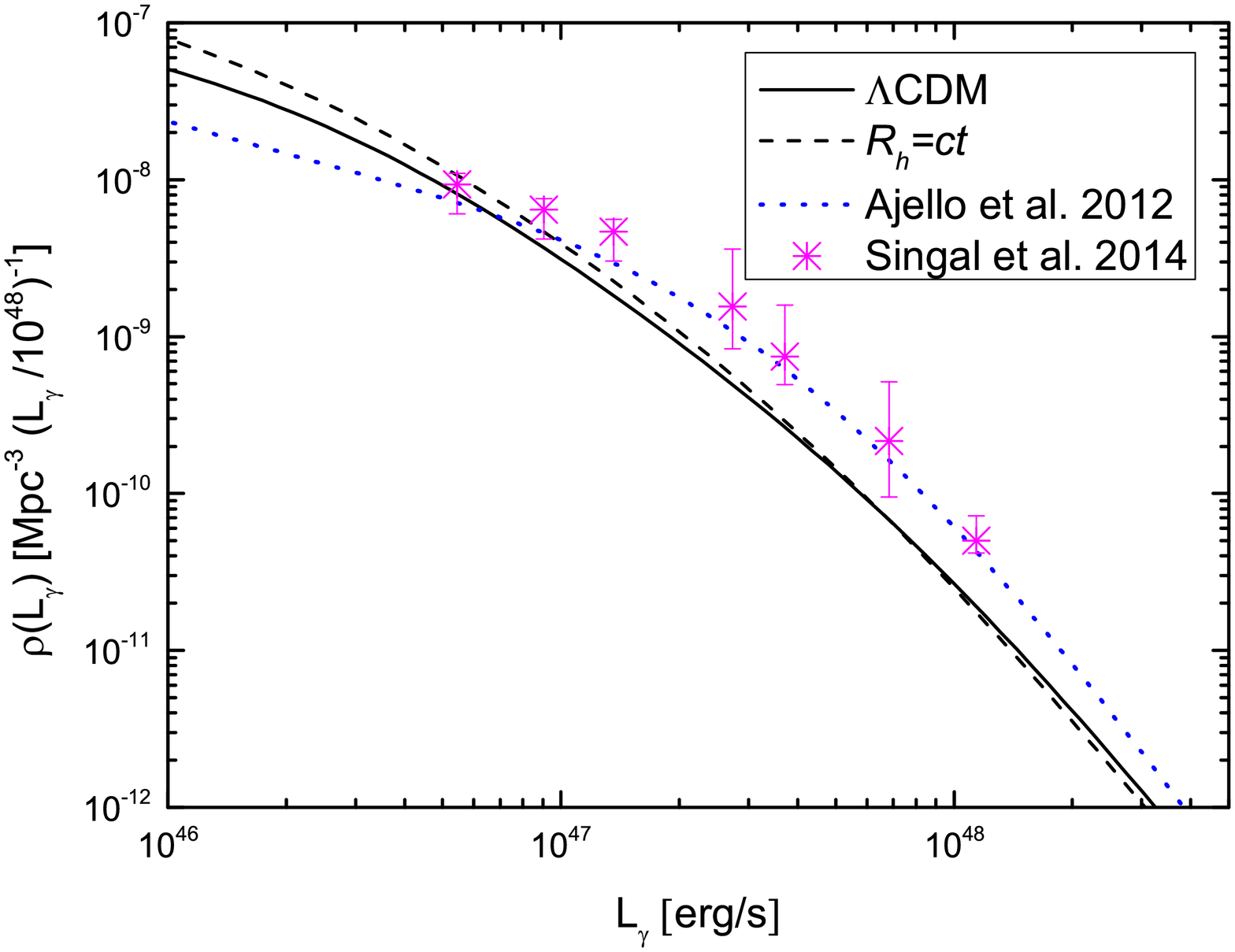}
    \end{tabular}
  \end{center}
    \caption{Differential local (left) and z=1 (right) $\gamma$-ray luminosity functions for
FSRQs assuming LDDE. The solid and dashed curves represent the results in this paper.
Stars are the results of \citet{Singal2014}, who restricted their analysis solely to FSRQs
with a $\gtrsim 7 \sigma$ detection threshold in the first-year catalog of
the \emph{Fermi} LAT. The dotted curves are the FSRQ LFs reported by \citet{Ajello12}.}
    \label{fig:figure10}
\end{figure*}

To complete our discussion, we also summarize here a comparison of our results
with others reported in the literature. Since LDDE appears to be strongly favoured
by the data over PLE, we will focus our attention on this particular GLF. Figure~\ref{fig:figure10}
compares the differential local (z=0) and z=1 GLFs with those reported by \cite{Ajello12}
and \cite{Singal2014}. \cite{Ajello12} analyzed the LF by using the sample comprised
of 186 FSRQs detected by \emph{Fermi} with $TS \geq 50$, $|b|\geq 15^{o}$ and
$F_{\gamma} \geq 10^{-8}$ photons cm$^{-2}$ s$^{-1}$. The results of \cite{Singal2014}
were obtained by analysing the sample of 184 FSRQs with $TS \geq 50$, $|b|\geq 20^{o}$
reported by Shaw et al. (2012). We can see that our distributions have a normalization
approximately two times larger than theirs. This is merely a reflection of the fact
that our sample (414) is about two times bigger than theirs (186 and 184); other than
this obvious difference, our results for the local universe are virtually identical to theirs. The right-hand
plot in Figure~\ref{fig:figure10} shows some slight differences in the determination of the GLF at $z=1$,
possibly due to the different redshift distributions of the various samples used
for the optimization of the model parameters or the incompleteness of the earlier samples. In
this regard, we note from the bottom panels of Figure~\ref{fig:figure9} that the observed data of our sample
are concordant with those of \cite{Ajello12} at high fluxes, but they clearly differ in the low flux
region. This would confirm the fact that our results should be the same at low redshifts, but
differ with \cite{Ajello12} at high redshifts, as is evident in Figure~\ref{fig:figure10}.


\subsection{Model Comparisons}

We now turn to the main goal of our analysis, which is to directly compare these two
cosmologies, for which we must use the model selection tools discussed in \S~3 above.
Starting with the PLE GLF (\S~5.1), the values of $W$ (from which $\Delta$ is calculated)
are shown in Table \ref{tab:table1}. Our optimization procedure shows that \textbf{$\Delta\equiv W_{\Lambda{\rm CDM}}
-W_{R_{\rm h}=ct}$} (for this assumed PLE luminosity function) is $\sim 134$, well into
the `very strong' category. At least for the PLE GLF, the size of our $\gamma$-ray
emitting FSRQ sample is already large enough for this statistical assessment to
overwhelmingly favour $R_{\rm h}=ct$ over the concordance $\Lambda$CDM model.

As we have seen, the LDDE GLF is a significantly better match to the data than
PLE. Here too, the model selection tools very strongly favour $R_{\rm h}=ct$
over $\Lambda$CDM. In the case of LDDE, $\Delta\equiv W_{\Lambda{\rm CDM}}-
W_{R_{\rm h}=ct}=144$, which again is well into the `very  strong' category.
The choice of GLF does not appear to have much influence in deciding which of
these two cosmologies is favoured by the \emph{Fermi} FSRQ data. The sample
is already large enough for the observations to strongly prefer the differential
volume dependence on $z$ predicted by $R_{\rm h}=ct$ over that in $\Lambda$CDM.

\section{Conclusions}
The extensive, high-quality sample of $\gamma$-ray emitting FSRQs observed by \emph{Fermi}
has generated considerable interest in identifying the $\gamma$-ray luminosity function and
its evolution with cosmic time. The number density of such objects has changed considerably
during the expansion of the Universe, growing dramatically up to redshift $\sim 0.5-2.0$
and declining thereafter. Aside from the obvious benefits one may derive from better
understanding this evolution as it relates to supermassive black-hole growth and its
connection to the halos of host galaxies, its strong dependence on redshift all the way out
to $z\sim 3$ offers the alluring possibility of using it to test different cosmological models.

In this paper, we have introduced this concept by directly comparing two specific
expansion scenarios, chiefly to examine the viability of the method. To do so, we
have opted to use prior values for the model parameters themselves, and instead
focus on the optimization of the parameters characterizing the chosen ansatz for the
luminosity function.  In doing so, one may question whether the choice of GLF unduly
biases the fit for one model or the other. This is a legitimate concern, and considerable
work still needs to be carried out to ensure that one is not simply customizing the
GLF for each background cosmology.

For this reason, we have opted in this paper to use two different forms of the GLF,
one for pure luminosity evolution and the second for a luminosity-dependent
density evolution, even though earlier work had already established a preference
by the data for the latter over the former. We have found that selecting either of
these GLFs has no influence at all on the outcome of model comparison tools. In
both cases, information criteria, such as the AIC, KIC, and BIC, show quite
conclusively that the evolution of the GLF for FSRQs very strongly favours
$R_{\rm h}=ct$ over the concordance $\Lambda$CDM model.

Cosmic evolution is now studied using a diversity of observational data, including
high-$z$ quasars (Melia 2013a, 2014), Gamma-ray bursts (Wei et al. 2013), the use of cosmic
chronometers (Melia \& Maier 2013; Melia \& McClintock 2015), type Ia supernovae
(Wei et al. 2015) and, most recently,
an application of the Alcock-Paczy\'nski test using model-independent Baryon Acoustic
Oscillation (BAO) data \citep{Font2014,Delubac2015,Melia2015b}, among others.
The BAO measurements are particularly noteworthy because, with their
$\sim 4\%$ accuracy, they now rule out the standard model in favour of $R_{\rm h}=ct$
at better than the $99.34\%$ C.L.

In this paper, we have provided a compelling confirmation of these other results by
demonstrating that population studies, though featuring a strong evolution in redshift,
 may also be used to independently check the outcome of model comparisons based purely
on geometric considerations. We emphasize, however, that much work still needs
to be done to properly identify how to best characterize the number density
function for this type of analysis. This would be critically important in cases,
unlike $\Lambda$CDM and $R_{\rm h}=ct$,  where cosmological models are
so different that an appropriate common ansatz may be difficult to find.

\section*{Acknowledgements}

We thank the referee, Mattia Di Mauro, for a careful reading of
our manuscript, and for thoughtful comments that have led to an
improvemed presentation, including several clarifying descriptions
of the results. We acknowledge the use of COSRAYMC (Liu et al.
2012) adapted from the COSMOMC package (Lewis \& Bridle 2002).
FM is grateful to Amherst College for its support through a John Woodruff Simpson
Lectureship, and to Purple Mountain Observatory in Nanjing, China, for its hospitality
while part of this work was being carried out. LZ acknowledges partial funding
support by the National Natural Science Foundation of China (NSFC) under grant
No. 11433004. This work was partially supported by grant 2012T1J0011 from The
Chinese Academy of Sciences Visiting Professorships for Senior International Scientists,
and grant GDJ20120491013 from the Chinese State Administration of Foreign Experts
Affairs. This work is also supported by the Key Laboratory of Particle Astrophysics
of Yunnan Province (Grant 2015DG035).
This work is also partially supported by the Strategic Priority Research Program,
the Emergence of Cosmological Structures, of the Chinese Academy of Sciences,
Grant No. XDB09000000, and the NSFC grants 11173064, 11233001, and 11233008.

\begin{figure}
	\includegraphics[width=\columnwidth]{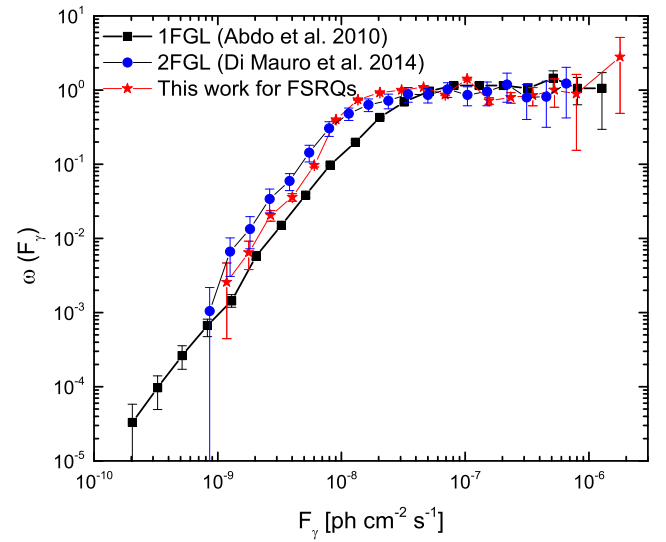}
    \caption{Efficiency for the sample of FSRQs evaluated using Eq. \ref{eq:efficieny}, in
    the region $\Gamma \in [1.5,3.2]$ for this work (indicated by red stars), and
    the detection efficiency of the 1FGL \citep{Abdo10c} and 2FGL \citep{DiMauro2014a} samples.}
    \label{fig:figure11}
\end{figure}









%


\bsp	
\label{lastpage}
\end{document}